\documentclass[aps,prc,twocolumn,superscriptaddress, showpacs]{revtex4-1}

\usepackage{graphicx}

\begin{document}

\title{ Branches of $^{33}$S($p,\gamma)^{34}$Cl at Oxygen-Neon Nova Temperatures}

\author{B. M. Freeman}
\email{blakef@u.washington.edu}
\affiliation{Department of Physics, University of Washington, Seattle, WA 98195-1560, USA}
\affiliation{CENPA, Box 354290, University of Washington, Seattle, WA 98195-4290, USA}
\author{C. Wrede}
\affiliation{Department of Physics, University of Washington, Seattle, WA 98195-1560, USA}
\affiliation{CENPA, Box 354290, University of Washington, Seattle, WA 98195-4290, USA}
\author{B. G. Delbridge}
\affiliation{Department of Physics, University of Washington, Seattle, WA 98195-1560, USA}
\affiliation{CENPA, Box 354290, University of Washington, Seattle, WA 98195-4290, USA}
\author{A. Garc\'{i}a}
\affiliation{Department of Physics, University of Washington, Seattle, WA 98195-1560, USA}
\affiliation{CENPA, Box 354290, University of Washington, Seattle, WA 98195-4290, USA}
\author{A. Knecht}
\affiliation{Department of Physics, University of Washington, Seattle, WA 98195-1560, USA}
\affiliation{CENPA, Box 354290, University of Washington, Seattle, WA 98195-4290, USA}
\author{A. Parikh}
\altaffiliation{Present address: Departament de F\'{i}sica i Enginyeria Nuclear, EUETIB, Universitat Polit\`{e}cnica de Catalunya, E-08036 Barcelona, Spain}
\affiliation{Physik Department E12, Technische Universit\"{a}t M\"{u}nchen, D-85748 Garching, Germany}
\affiliation{Maier-Leibnitz-Laboratorium der M\"{u}nchner Universit\"{a}ten (MLL), D-85748 Garching, Germany}

\author{A. L. Sallaska}
\affiliation{Department of Physics, University of Washington, Seattle, WA 98195-1560, USA}
\affiliation{CENPA, Box 354290, University of Washington, Seattle, WA 98195-4290, USA}

\date{\today}

\begin{abstract}
Recent simulations of classical novae on oxygen-neon white-dwarf stars indicate that the isotopic ratio $^{32}$S/$^{33}$S  has the potential to be a remarkable indicator of presolar grains of nova origin. The $^{33}$S$(p,\gamma)^{34}$Cl reaction influences this ratio directly by destroying $^{33}$S in novae. Additionally, $\beta$ delayed $\gamma$-rays from the metastable state of $^{34}$Cl ($t_{1/2}$ = 32~min) have been suggested to be potential nova observables. We have measured the branches for known  $^{33}$S$(p,\gamma)^{34}$Cl resonances that are activated at temperatures relevant to oxygen-neon novae. We provide the first reliable uncertainties on these branches and the first upper limits for several previously unmeasured branches.  

\end{abstract}

\maketitle

\emph{Introduction.} Classical novae explosions arise from thermonuclear runaways on white dwarf stars, resulting from the accretion of hydrogen-rich gas from a companion star in a binary system. Among sites of explosive nucleosynthesis, novae are currently of particular astrophysical interest because their peak temperatures, $T_{peak} = 0.1-0.4$~GK~\cite{Jose2001}, are low enough that they may be modeled using thermonuclear reaction rates that are based mostly on experimental data~\cite{Iliadis2002}. These models may be compared to observations through direct spectroscopy of nova ejecta, isotopic measurements of presolar grains, and cosmic $\gamma$-ray emitters.

	Presolar grains are micron-sized grains of material, found in primitive meteorites, identified through isotopic ratios that differ from those in the solar system at large. Although most grains are believed to have originated from supernovae and asymptotic giant branch (AGB) stars~\cite{Lugaro2005}, several grains have recently been identified as having low $^{12}$C/$^{13}$C, $^{14}$N/$^{15}$N and high $^{30}$Si/$^{28}$Si isotopic ratios relative to terrestrial values. These ratios indicate a possible classical nova origin~\cite{Amari2001, Nittler2005}, but additional isotopic signatures would make these identifications more robust. Models of classical novae on a 1.35~$M_{\odot}$ oxygen-neon (ONe) white dwarf yield a $^{32}$S/$^{33}$S abudance ratio between 30 and 9700~\cite{Parikh2009} that may be compared to the solar ratio of 127. Therefore, anomalous $^{32}$S/$^{33}$S ratios have the potential to provide a clear signature for presolar grains of ONe nova origin~\cite{Jose2004, Amari2001}. However, the predicted ratio is subject to large uncertainties owing to insufficient experimental information on the $^{33}$S($p,\gamma)^{34}$Cl reaction at nova temperatures.

 	The $^{33}$S($p,\gamma)^{34}$Cl reaction is also relevant to $\gamma$-ray astronomy because it produces $^{34}$Cl in ONe novae. $^{34}$Cl has an isomeric state ($^{34}$Cl$^{m}$) located at $E_{x}$ = 147~keV with $t_{1/2}$ = 32~min~\cite{Endt1990}. $\beta$-delayed $\gamma$-rays from $^{34}$Cl$^{m}$ of energy 1.177, 2.128, 3.304~MeV, have been suggested as possible nova observables~\cite{Leising1987, Coc2000}. However, several factors may contribute to a considerable suppression of external $^{34}$Cl$^{m}$ $\beta$-delayed $\gamma$-ray flux~\cite{Parikh2009}. These factors include the short half life of $^{34}$Cl$^{m}$ combined with predictions that the half-life is effectively reduced by thermally induced transitions to the ground state~\cite{Coc2000} and that the envelope of an ONe nova remains optically dense for hours to days following $T_{peak}$~\cite{Gomez-Gomar1998, Hernanz1999, Jose2007}. Resonances in the  $^{33}$S($p,\gamma)^{34}$Cl  reaction at energies corresponding to the temperature region of interest for novae, and their $\gamma$-ray branching ratios, play a role in determining the population of $^{34}$Cl$^{m}$ and, hence, its viability as a nova observable. Although detection of such $\gamma$-rays from a nova outburst seems highly unlikely,  better knowledge of the $^{33}$S$(p,\gamma$) rate and the corresponding $\gamma$-ray branches from proton-capture resonances in $^{34}$Cl is needed to make firm predictions of the expected $\gamma$-ray flux.

	 The $^{33}$S($p,\gamma)^{34}$Cl reaction has been measured previously in the region $E_{r}$ = 0.4 - 2.0~MeV yielding resonance strengths, spins and parities of excited states~\cite{Dassie1, Dassie2, Glaudemans1964, Waanders1983}. However, prior to the present experiment, the only measurements of $\gamma$-decay schemes for the $^{33}$S($p,\gamma)^{34}$Cl resonances that dominate the ONe nova reaction rate ($E_{r} = 432, 492, 529$~keV) were provided by Glaudemans \textit{et al.}~\cite{Glaudemans1964} and Waanders \textit{et al}~\cite{Waanders1983}. The former was performed with NaI scintillators with relatively poor energy resolution and was interpreted without the benefit of current knowledge of the bound-level structure of $^{34}$Cl. The latter was performed with Ge(Li) detectors with good energy resolution but was focused on branches of bound levels; uncertainties for the branches of the measured resonances were unfortunately not reported. 
	 
	 In this Brief Report, we present measurements of $\gamma$-ray branches for these $^{33}$S($p,\gamma)^{34}$Cl resonances and provide reliable uncertainties for the first time. Our measurements will complement those from an experiment using the DRAGON facility~\cite{Engel2005, Hutcheon2003} at TRIUMF-ISAC that seeks to determine the strengths of  $^{33}$S($p,\gamma)^{34}$Cl  resonances in the temperature range of interest for novae~\cite{ParikhTRIUMF}. The TRIUMF-ISAC experiment uses inverse kinematics with a recoil separator and bismuth germanate (BGO) $\gamma$-ray detectors surrounding the target. Owing to the relatively poor energy resolution of BGO detectors, it is difficult to measure branching ratios using DRAGON; also the efficiency of the BGO array depends on the $\gamma$-decay scheme. Therefore, we undertook an independent experiment for this purpose.

	\emph{Experiment.} The $\gamma$-ray branching measurements were performed at the Center for Experimental Nuclear Physics and Astrophysics at the University of Washington with an existing setup previously used to measure the $^{22}$Na($p,\gamma)^{23}$Mg reaction~\cite{Anneprc, Anneprl, Annethesis}. The setup consisted of a beam line with a target chamber and two high-purity germanium detectors at $\theta_{lab} = \pm$~55$^{\circ}$. This detector geometry was chosen to coincide with zeroes of the Legendre polynomial P$_{2}$ in the laboratory frame, with the intention of reducing systematic errors associated with $\gamma$-ray angular distributions. The detectors were surrounded by plastic scintillators used to reduce cosmic ray background via anti-coincidence and by passive lead shielding. A tandem Van de Graaff accelerator in terminal-ion-source mode was used to accelerate protons to lab energies between 200 and 710 keV with currents of $\sim$~45~$\mu$A. The only major change to the experimental setup following the previous experiment was the removal of 26-mm of lead shielding between the target and detector system. An effort was also made to clean up residual $^{22}$Na in the beamline from the previous experiment to minimize $\gamma$-ray background from its $\beta$ decay.

	Targets were prepared by implanting $\sim$~10$^{16}$ $^{33}$S ions into rectangular oxygen-free high-conductivity copper substrates at CENPA. Using a General Ionex 860 sputter ion source, a 45-keV $^{33}$S beam was produced, isolated with a 90$^{\circ}$ analyzing magnet, then rastered over a circular 5-mm-diameter collimator positioned 5 cm upstream of the copper substrate. The methods and apparatus used for target preparation were similar to those used for $^{23}$Na in Ref.~\cite{Brown2009}.
	
\begin{figure}

\includegraphics[width=.50\textwidth]{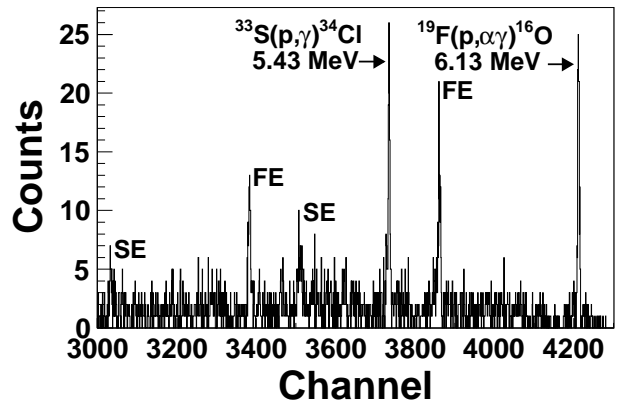}

\caption{
$^{33}$S($p , \gamma)^{34}$Cl spectrum from the  $E_r = 432$~keV resonance showing the 5426~keV line corresponding to r~$\rightarrow$~147~keV as well as the 6130~keV background peak from the  $^{19}$F($p , \alpha \gamma)^{16}$O reaction. Photopeaks, first escape peaks (FE) and second escape peaks (SE) are labeled.
}

\label{33Sfig1} 

\end{figure}

	\emph{Data and Analysis.}  Efficiencies were determined in the $^{22}$Na$(p,\gamma)$ experiment using a combination of PENELOPE Monte-Carlo simulations, measurements of a calibrated $^{60}$Co source, $^{24}$Na sources, and $^{27}$Al$(p,\gamma)$ reaction branches~\cite{Anneprc}.  Corrections to account for the removal of the lead shielding were applied. As in Ref.~\cite{Anneprc} an additional $\pm$~3\% systematic uncertainty was assigned to account for angular-distribution effects.

	 The $\gamma$-ray energy calibration was performed using the $^{40}$K (1.461~MeV) and $^{209}$Tl (2.615~MeV) background peaks, as well as the 6.130~MeV $\gamma$-ray from the  $^{19}$F($p , \alpha \gamma)^{16}$O reaction and the 1.275~MeV $\gamma$-ray from $^{22}$Na. Corrections for recoil losses and Doppler shift were applied. All calibration peaks  were fit using a Gaussian function, and the resulting centroid channels were fit linearly versus energy. At 1275~keV our detectors have energy resolutions of 3.7, and 3.9~keV.  A partial $\gamma$-ray energy spectrum for the $E_{r}$ = 432~keV resonance is shown in Fig. \ref{33Sfig1}. A linear background subtraction was employed in the integration of measured peaks.

	The most significant backgrounds were from $^{22}$Na decay ($t_{1/2}$ = 2.603~years, $E_{\gamma}$ = 1.275~MeV) and the $^{19}$F($p,\alpha \gamma)^{16}$O reaction. Additionally, at $E_{r}$ = 529~keV
 we observed background from a strong $^{13}$C($p,\gamma)^{14}$N resonance ($E_{r}$ = 558~keV, $\omega\gamma$ =  8.8~eV). In some cases these factors were severe enough to prevent measurement of particular branches, especially at $\gamma$-ray energies below $\sim$~2.4~MeV. Due to the strength of the 1.275~MeV peak and the related pileup, it was not possible to measure any branches with $E_{\gamma}~\lesssim~$1.8~MeV.

\begin{table*}
\caption{Relative intensities of primary $\gamma$-ray branches for measured $^{33}$S($p,\gamma)^{34}$Cl (Q = 5142.5 $\pm$ 0.2 keV~\cite{Waanders1983}) resonances normalized to the strongest observed branch. Uncertainties associated with the strongest branch (7.4$\%$, 3.5$\%$, 6.7$\%$ at $E_r$~=~432, 492, 529~keV respectively) were accounted for in other branches. Comparisons with results from Waanders \textit{et al.}~\cite{Waanders1983} are included.  \label{t1}}
\begin{ruledtabular}
\begin{tabular}{ c| c c c c c c }	

Final $^{34}$Cl &	\multicolumn{2}{c}{$E_r = 432$ keV}  &  	\multicolumn{2}{c}{$E_r = 492$ keV}  &	 \multicolumn{2}{c}{$E_r = 529$ keV}  \\
\cline{2-3} \cline{4-5} \cline{6-7}
level (keV) & Present & Waanders \textit{et al.}~\cite{Waanders1983} & Present & Waanders \textit{et al.}~\cite{Waanders1983} & Present & Waanders \textit{et al.}~\cite{Waanders1983}\\
\hline

0 	&				&	$< 0.2$	&100		&	100		&100		&	100	\\

147 	&61.9$\pm$5.5		&	61.9		&	$<0.65$		&	$<0.4$	&	$<1.4$		&	$<0.4$\\

461	&				&	$< 1.4$	&2.2$\pm$0.7		&	2.9		&	$<3.8$		&	$<0.5$\\

666	&				&	$<0.7$	&15.4$\pm$1.0		&	15.9		&6.4$\pm$3.4		&	8.2\\

1230	&10.3$\pm$1.4		&	10		&0.6$\pm$0.4		&	0.6		&				&		\\

1887&5.5$\pm$1.0		&	4.8		&1.8$\pm$0.3		&	1.9		&	$<4.4$		&	1.8	\\

2158&				&			&0.6$\pm$0.3		&			&8.8$\pm$2.4		&	10 \\

2181&				&			&	$<0.5$		&			&	$<3.2$		&		\\

2376&3.0$\pm$1.2		&	1.0		&	$<0.3$		&			&	$<3.0$		&		\\

2580&	$<1.2$		&			&12.2$\pm$0.7		&	14.5		&11.6$\pm$2.6		&	6.6\\

2611&				&	0.7		&	$<0.4$		&			&	$<1.5$		&		\\

2721&22.6$\pm$2.3		&	21.2		&1.0$\pm$0.3		&	0.7		&	$<3.8$		&		\\

3129&	$<3.9$		&			& 3.9$\pm$0.4		&	4.1		&				&	2.8	\\

3334&				&			& 				&			&				&		\\

3383&				&			&				&			&	$<1.3$		&		\\

3545&100		&	100		&				&			&				&		\\

3600&28.7$\pm$6.6		&	38.1		&				&			&	$<5.5$		&		\\

3632&				&			&	$<0.3$		&			&	$<4.0$		&		\\

3646&				&			&	$<0.4$		&			&	$<3.3$		&		\\

3660&				&			&	$<0.3$		&			&	$<3.4$		&		\\

3774&				&			&				&			&	$<4.6$		&	0.5	\\

3792&				&			&3.7$\pm$0.7		&	2.5		&	$<6.7$		&	1.9 \\

\end{tabular}	
\end{ruledtabular}
\end{table*}

	\emph{Results and Discussion.} Branches are shown in Table~I, normalized to the strongest branch for each resonance, including a comparison with the results from Waanders \textit{et al.}~\cite{Waanders1983}. We were able to provide upper limits for several transitions that were not reported in Ref.~\cite{Waanders1983}. 
	
	%The branches not reported by Waanders \textit{et al.}~\cite{Waanders1983} or the present authors are likely due to $^{19}$F($p,\alpha \gamma)^{16}$O background common to both experiments. Although it is not discussed in Ref.~\cite{Waanders1983} it is possible that $^{13}$C($p,\gamma)^{14}$N background peaks at the E$_{r}$~=~529~keV resonance interfered with the measurement of several branches that were not reported by either the present work or Ref.~\cite{Waanders1983}.

	The present branches and those of Ref.~\cite{Waanders1983} are in good agreement, most importantly at the $E_{r}$ = 432~keV resonance, which could be the most significant contributor to $^{33}$S destruction at nova temperatures~\cite{Jose2007}. Figure~2 shows the normalized differences between our values and those of Ref.~\cite{Waanders1983}. The fact that the distribution in Fig.~2 agrees well with Gaussian statistics and is not broadened indicates that the unreported uncertainties in the data of Ref.~\cite{Waanders1983} are likely relatively small. Most importantly, regarding the relative production of $^{34}$Cl$^{m}$ versus $^{34}$Cl$^{g}$, our $E_{r}$ = 432~keV branch r~$\rightarrow$~147~keV agrees very well with Ref.~\cite{Waanders1983}. We do not make direct comparisons with the data from Glaudemans \textit{et al.}~\cite{Glaudemans1964}  because their limited knowledge of the bound states of $^{34}$Cl may have led to misinterpretations of the decay scheme. However, in the important case of the $E_{r}$ = 432~keV branch r~$\rightarrow$~147~keV, if we assume that they could not resolve r~$\rightarrow$~3545 keV and r~$\rightarrow$~3600~keV branches, by summing our measurements of these branches and renormalizing we find agreement between our r~$\rightarrow$~147~keV branch and the branch in Ref~\cite{Glaudemans1964}.
\begin{figure}[b]

\includegraphics[width=.50\textwidth]{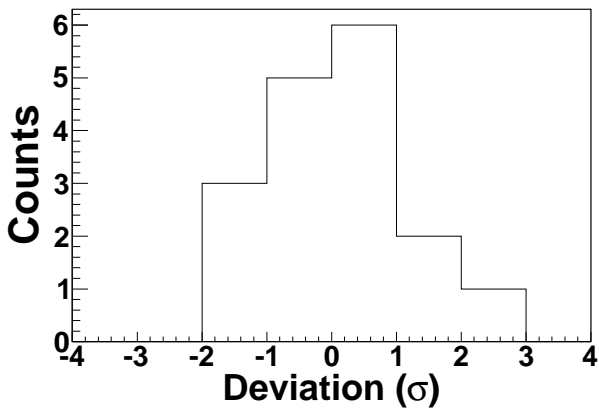}

\caption{
Differences between relative $\gamma$-ray intensities for primary $^{33}$S$(p,\gamma)$ branches reported by Waanders \textit{et al.}~\cite{Waanders1983} and those in the present experiment.
}

\label{sigma} 

\end{figure}	

\begin{figure}[b]

\includegraphics[height=.50\textwidth]{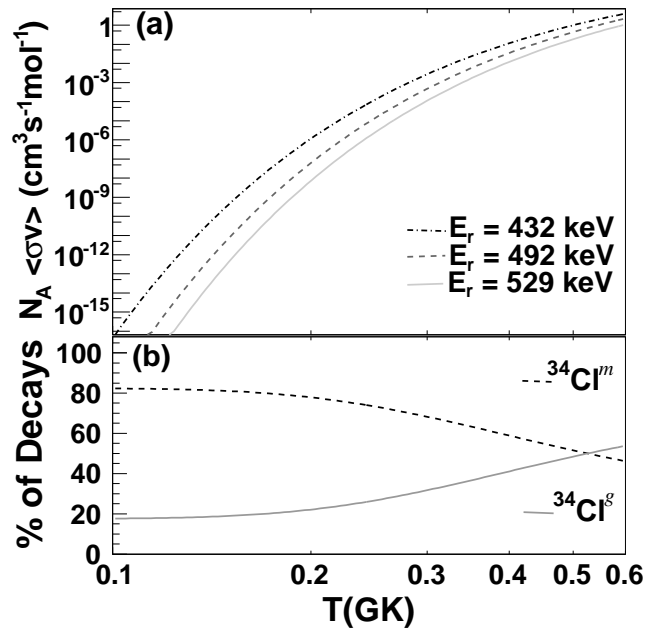}

\caption{ \label{rate}
(a): Stellar $^{33}$S($p,\gamma)^{34}$Cl reaction rate for each measured resonance, using strengths from Ref.~\cite{Waanders1983}. (b): Calculated percentage of decays resulting in $^{34}$Cl$^{g}$ versus $^{34}$Cl$^{m}$ using data from this work.
}

\end{figure}

	Figure \ref{rate} (a) shows the stellar $^{33}$S($p,\gamma)^{34}$Cl reaction rate for measured resonances calculated using resonance strengths from Ref.~\cite{Waanders1983}. Using the finite primary branches from our data combined with branchings for bound levels from Ref.~\cite{Waanders1983}, we calculate the percentage of decays from each measured resonance that ends in the $^{34}$Cl$^{m}$ state: $82.5 \pm 6.9\%$ for $E_{r}$ = 432~keV, $2.2 \pm 1.3\%$ for $E_{r}$ = 492~keV, and $1.0 \pm 0.3\%$ for $E_{r}$~=~529~keV. Using these central values combined with resonance strengths from Ref.~\cite{Waanders1983} we calculate the total percentage of decays producing  $^{34}$Cl$^{m}$ and $^{34}$Cl$^{g}$ and plot them versus temperature in Fig. \ref{rate} (b). Given that the $E_{r} = 432$ keV resonance dominates the reaction rate between $T_{peak} = 0.1-0.4$~GK, we find mostly $^{34}$Cl$^{m}$ produced in this range.

	We also acquired data on potential resonances at $E_r = 214$, 244, 260, 281, 301, 342, and 399~keV that were discovered or compiled in Refs.~\cite{Parikh2009, Parikhprep}; these potential resonances have never been observed through direct $^{33}$S($p$,$\gamma)^{34}$Cl measurements. Each potential resonance was measured for roughly 6.5 to 23~h using beam energies that were expected to maximize the yields (full excitation functions were not measured). We did not observe statistically significant $^{33}$S($p$,$\gamma)^{34}$Cl yields at any of these energies. Based on the non-observation of these resonances and statistical considerations alone we would be able to set upper limits of $\approx 1$~meV or lower on the partial strengths of any branches from these resonances with primary $\gamma$-ray energies above our effective detection threshold of 2.4~MeV, provided they were not obscured by $^{19}$F($p$,$\alpha\gamma$)$^{16}$O background peaks. However, our measurement was not planned or conducted to set absolute limits on resonance strengths and therefore large systematic uncertainties associated with target stability, beam-target alignment, and optimal beam energy must also be considered. Taking into account very conservative estimates of these uncertainties, we find it to be highly unlikely that any branches from these potential $^{33}$S($p,\gamma)^{34}$Cl resonances below $E_r = 400$~keV have partial resonance strengths greater than 10~meV, subject to the conditions on threshold and background already stated. Assuming the potential resonance at 342~keV is dominated by one or two branches that would have been observable in our experiment, our limits improve upon the estimate of $\omega\gamma <$ 65~meV from Ref.~\cite{Parikh2009}, reducing the potential contribution of this resonance to the thermonuclear $^{33}$S($p$,$\gamma)^{34}$Cl reaction rate substantially.

	\emph{Conclusion.} We have provided the first measurements of  branches for $^{33}$S$(p,\gamma)^{34}$Cl resonances that include reliable uncertainties and upper limits for 18 previously unmeasured branches. Our results contribute essential information concerning the production of sulfur isotopes and $^{34}$Cl$^{m}$ in novae. Additionally, our results will complement the experiment at TRIUMF-ISAC~\cite{ParikhTRIUMF}.

We would like to thank G. Harper and D.I. Will for their assistance in the preparation of implanted targets, and in the operation of the tandem Van de Graaff accelerator. This work was supported by the United States Department of Energy under contract DE-FG02-97ER41020. AP was supported by the DFG cluster of excellence ``Origin and Structure of the Universe" (www.universe-cluster.de).

\end{document}